\documentclass[fleqn,10pt]{wlscirep}
\usepackage[utf8]{inputenc}
\usepackage[T1]{fontenc}

\usepackage{amsmath,amssymb,amsfonts}
\usepackage{algorithmic}
\usepackage{graphicx}
\usepackage{textcomp}
\usepackage{cite}
\usepackage{caption}
\usepackage{subcaption}
\usepackage[export]{adjustbox}
\usepackage{multirow}
\usepackage{url}
\usepackage{makecell}
\usepackage{float}
\usepackage[capitalise,noabbrev]{cleveref}
\usepackage{xr}

\makeatletter
\newcommand*{\addFileDependency}[1]{% argument=file name and extension
  \typeout{(#1)}
  \@addtofilelist{#1}
  \IfFileExists{#1}{}{\typeout{No file #1.}}
}
\makeatother

\title{Reproducibility analysis of automated deep learning based localisation of mandibular canals on a temporal CBCT dataset}

\author[1,2]{Jorma J\"arnstedt}
\author[3]{Jaakko Sahlsten}
\author[3]{Joel Jaskari}
\author[3,6,*]{Kimmo Kaski}
\author[1]{Helena Mehtonen}
\author[4]{Ari Hietanen}
\author[4]{Osku Sundqvist}
\author[4]{Vesa Varjonen}
\author[4]{Vesa Mattila}
\author[5]{Sangsom Prapayasotok}
\author[5]{Sakarat Nalampang}

\affil[1]{Medical Imaging Centre, Department of Radiology Tampere University Hospital, Teiskontie 35, 33520 Tampere, Finland}
\affil[2]{The Graduate School, Chiang Mai University, 239 Huaykaew Road, Suthep, Meuang, Chiang Mai, Thailand}
\affil[3]{Aalto University School of Science, Otakaari 1, 02150 Aalto, Finland}
\affil[4]{Planmeca Oy, Asentajankatu 6, 00880 Helsinki, Finland}
\affil[5]{Division of Oral and Maxillofacial Radiology, Faculty of Dentistry, Chiang Mai University, Suthep Rd., T. Suthep, A. Muang, Chiang Mai, Thailand}
\affil[6]{Alan Turing Institute, British Library, 96 Euston Rd, London NW1 2DB, UK}
\affil[*]{Corresponding author, kimmo.kaski@aalto.fi}


\begin{abstract}

Preoperative radiological identification of mandibular canals is essential for maxillofacial surgery. This study demonstrates the reproducibility of a deep learning system (DLS) by evaluating its localisation performance on 165 heterogeneous cone beam computed tomography (CBCT) scans from 72 patients in comparison to an experienced radiologist's annotations. We evaluated the performance of the DLS using the symmetric mean curve distance (SMCD), the average symmetric surface distance (ASSD), and the Dice similarity coefficient (DSC). The reproducibility of the SMCD was assessed using the within-subject coefficient of repeatability (RC). Three other experts rated the diagnostic validity twice using a 0-4 Likert scale. The reproducibility of the Likert scoring was assessed using the repeatability measure (RM). The RC of SMCD was 0.969 mm, the median (interquartile range) SMCD and ASSD were 0.643 (0.186) mm and 0.351 (0.135) mm, respectively, and the mean (standard deviation) DSC was 0.548 (0.138). The DLS performance was most affected by postoperative changes. The RM of the Likert scoring was 0.923 for the radiologist and 0.877 for the DLS. The mean (standard deviation) Likert score was 3.94 (0.27) for the radiologist and 3.84 (0.65) for the DLS. The DLS demonstrated proficient qualitative and quantitative reproducibility, temporal generalisability, and clinical validity.
\end{abstract}
\begin{document}

\flushbottom
\maketitle

\thispagestyle{empty}

\section*{Introduction}\label{sec:introduction}
The identification of bilateral mandibular canals from preoperative radiological examinations is important before oral and maxillofacial surgical operations in the lower jaw area, such as implantology, impacted teeth operative extraction, and pathological lesion resection or enucleation~\cite{agbaje2017tracking}. The mandibular canals are seen as radiolucent structures surrounded by cortical bony borders with two openings; the foramen mentale in the anterior and foramen mandibulae in the posterior part. The thickness, location, and shape of the mandibular canal can have a number of anatomical variations, and some of them are impossible to identify due to difficult bone quality and pathological changes~\cite{bertl2014histomorphometric, oliveira2011visibility}. These canals host a neurovascular bundle, including an artery, veins, and the inferior alveolar nerve that supplies motoric and sensory innervations to chin, lower lips, and teeth. Damaging the mandibular canals may cause excessive bleeding, and temporary or permanent neurological complications~\cite{rood1990radiological}.
The challenging and labour intensive localisation of these canals is currently done manually, which poses a challenge as there are not enough specialists to meet the diagnostic needs~\cite{renard2020variability}. Automated tools show promise to alleviate the manual burden~\cite{brown2014basic, macleod2008cone}.

In order to locate the mandibular canals accurately, three dimensional (3D) radiological imaging is the recommended approach. For this purpose, Cone Beam Computed Tomography (CBCT) is commonly used~\cite{liang2010comparative}. However, there are diagnostic challenges due to technical and patient related artefacts, as well as heterogeneities.

In 2020, the International Society for Strategic Studies in Radiology published recommendations for technical validation in radiological AI research with four crucial properties: accuracy, robustness, generalisability, and reproducibility. The accuracy provides insight to the expected performance, robustness to the invariance to small pertubations in the data, and generalisability to the behaviour with various medical imaging  variabilities~\cite{recht2020integrating}. The reproducibility is defined as invariance to measurement variability on the same subject under changing conditions and can be evaluated from repetitive scans~\cite{sunoqrot2021reproducibility}. There are several reasons for repetitive CBCT scanning of patients, such as pathological changes, operation planning, and treatment follow-up~\cite{braun2022dental}. Without evaluating reproducibility, there is no evidence that a same image derived feature or element could be used in a longitudinal assessment~\cite{recht2020integrating}. 

Imaging-based deep learning models perform computations on the voxels of a scan. There are always at least slight changes in the voxel intensities when a patient is scanned again using the same device and imaging parameters at different times. These changes alter the internal computations of the deep learning models, but the effect on the model output should be negligible in the case of a robust and reproducible model. Therefore, there is a close relationship between the reproducibility and the overall performance of the model~\cite{kim2020test}. 

Recently, multiple deep learning based automatic mandibular canal segmentation systems have been introduced~\cite{jaskari2020deep, cipriano2022deep, issa2022effectiveness, jarnstedt2022comparison}. These studies have reported on accuracy and robustness with good performance, with two of them also reported the effect of heterogeneities on the deep learning performance~\cite{jaskari2020deep, jarnstedt2022comparison} and one of them the deep learning generalisability as well as interobserver performance~\cite{jarnstedt2022comparison}. However, none of these works have analysed the reproducibility of the deep learning systems.

The aim of this study is to validate the reproducibility and temporal generalisation of a previously introduced deep learning fully convolutional neural network-based automatic mandibular canal localisation system~\cite{jaskari2020deep, jarnstedt2022comparison} using both quantitative and qualitative evaluation.

\section*{Methods}\label{sec:methods}
In this section, we present a detailed description of the deep learning system and its components, describe the patient data used for the results, and the quantitative and qualitative evaluation measures we use to evaluate reproducibility.

\subsection*{Deep learning model}
The deep learning system (DLS) we study was proposed in two recent studies~\cite{jaskari2020deep,jarnstedt2022comparison}. This model was also used as a basis in the work of publicly available voxel level annotated 3D mandibular canal dataset~\cite{cipriano2022deep}. The DLS consists of two algorithms; a deep learning neural network that segments the voxels containing the mandibular canals from 3D CBCT volumes and a postprocessing algorithm that extracts the two most likely mandibular canal curves from the segmentation volume. For our analysis, we utilise the trained model from~\cite{jarnstedt2022comparison} to evaluate its temporal generalisability and reproducibility with a new dataset.

The deep learning model of DLS is a type of fully convolutional neural network with a U-net style architecture~\cite{ronneberger2015u}. The model utilises three dimensional convolutional layers that enables the recognition of patterns simultaneously in the axial, coronal, and sagittal planes. The deep learning model produces approximate segmentations of the mandibular canals, which are then postprocessed using connected component analysis for the purpose of reducing false positives and connecting partial segmentations to form a single canal. Segmentation is then reduced into a curve using a skeletonisation routine~\cite{lee1994building} and a heuristic concatenation algorithm is used to give as an output the mandibular canals based on correct anatomical characteristics and symmetry. 

\subsection*{Patient data}
The data was collected retrospectively from the Tampere University Hospital (TAUH) in 
Tampere Finland. The patients whose data had been previously used in the development, validation, or testing of the DLS were excluded. Every patient has at least two CBCT scans imaged for medical reasons. These reasons include benign lesions follow-up, operation planning and control for orthognathic surgery and temporomandibular joint prosthesis, trauma controls, computer assisted surgery planning, and bone destructive lesions including malignancies. The collected dataset of 165 CBCT scans consisted of 72 patients [mean age 43.1 years] with 47\% males, [mean age 42.5] and 53\% females [mean age 43.5], The scans were pseudonymised and categorised into four main subgroups to evaluate the effects of different heterogeneities. In addition, the dataset included immediate postoperative scans that were categorised into separate heterogeneity subgroups, to see the effect of the surgical operation before the healing process.

Three scanners with different imaging parameters were used i.e.\ KAVO OP 3D PRO, VISOG7 and Scanora 3Dx with 7, 66 and 92 scans, respectively. The scans included a range of voxel spacing from 0.2~mm to 0.5~mm with most the most common ones being 0.2~mm, 0.3~mm, 0.4~mm with 57\%, 29\% and 7\% of the scans, respectively. The number of CBCT scans with different devices and voxel spacings are presented in Supplementary Table S1. All the scans were whole facial imaging, large field-of-view scans, due to the aforementioned diagnostic needs in radiological imaging. There were 39 patients with two CBCT scans, and 33 patients with multiple (three to five) scans in the study material. There were 11 patients who were scanned with the same CBCT device and the same imaging parameters, 16 scanned with the same device but different imaging parameters, and 45 who were scanned with different devices and different imaging parameters. All the scans were taken in TAUH from 2015 to 2020.

\begin{figure*}[ht]%
\centering
\includegraphics[width=0.9\textwidth]{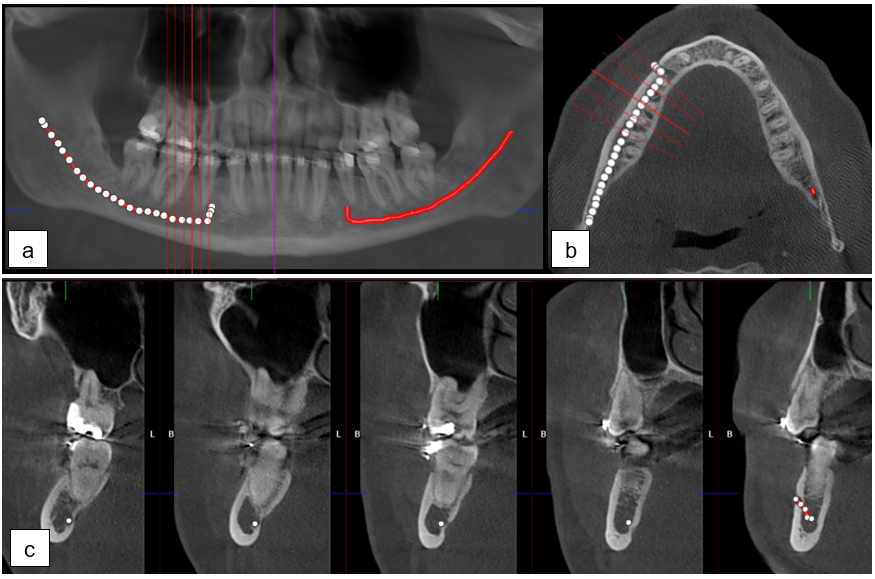}
\caption{ a.\ CBCT panoramic reconstruction showing the annotation control points in the left and the interpolated mandibular canal in right side of the mandible, b. an axial slice with cross-sectional slices along the panoramic curve shown in red, c.\ annotation control points in the cross-sectional slices 3~mm apart with multiple annotation points in the foramen mentale area}\label{fig1}
\end{figure*}

The mandibular canals were annotated using clinical software: Romexis 4.6.2, of Planmeca, Finland, by a senior radiologist with over 35 years of experience in dentistry, similarly to previous works~\cite{jaskari2020deep, jarnstedt2022comparison}. In the annotation process, a CBCT panoramic reconstruction was made (Fig.~\ref{fig1}a), cross-sectional images against panoramic curves were reconstructed (Fig.~\ref{fig1}b), and the midline of the canals were marked with standard 1.5~mm diameter control points placed in 3~mm intervals. Multiple control points were used for each of the canals, especially in the foramen mentale curvature area and in complicated cases (Fig.~\ref{fig1}c). An approximate segmentation was created when computing segmentation measures, in which the path of the canal was expanded to 1.5~mm static diameter tube.

The four main subgroups manifest various clinical and operative situations in dental and maxillofacial radiology, including pre- and postoperative, and follow-up CBCT scans. These subgroups were determined as \textit{Normal group}, \textit{Prosthetic group}, \textit{Orthognathic surgery group}, and \textit{Pathological group} with 23, 13, 22, 14 patients and 51, 35, 46, 33 scans, respectively.

\begin{figure*}[ht]%
\centering
\includegraphics[width=0.9\textwidth]{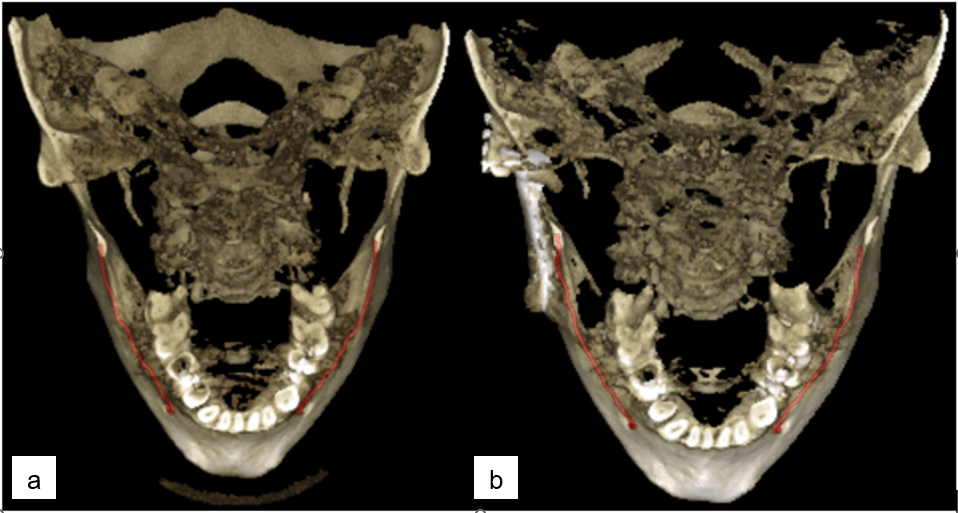}
\caption{Three dimensional reconstruction images with manually annotated mandibular canals shown in red, a.\ preoperative CBCT scan, b.\ postoperative CBCT scan with right side temporomandibular joint reconstruction on the angulus area of the mandible}\label{fig2}
\end{figure*}

\begin{figure*}[ht]%
\centering
\includegraphics[width=0.9\textwidth]{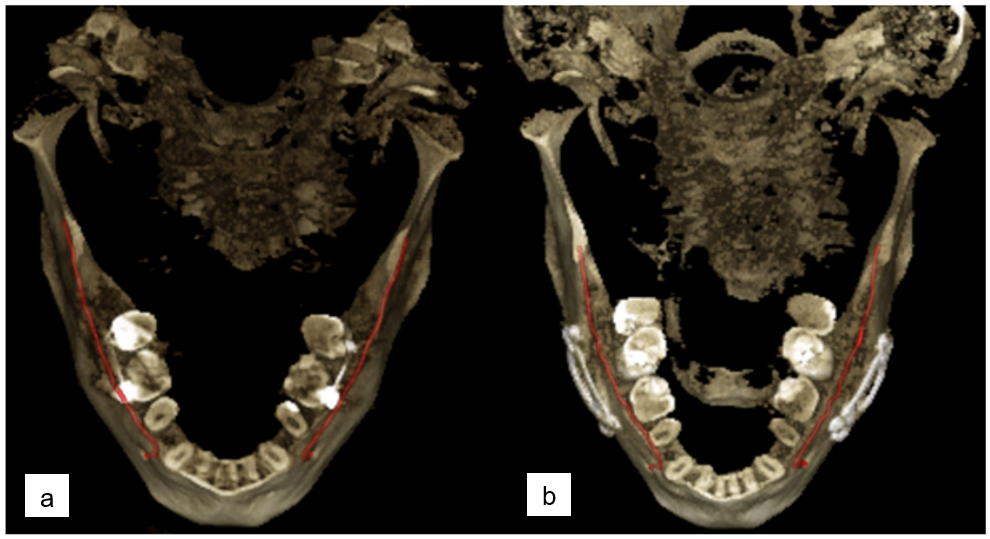}
\caption{Three dimensional reconstruction images of an orthognathic surgery patient showing changes in the postoperative scan with manually annotated mandibular canals shown in red, a. preoperative CBCT scan, b. postoperative CBCT scan 1 year after the operation}\label{fig3}
\end{figure*}

The patients of Normal group were scanned for reasons having no major influence in the mandibular canal area, but 14 of the 51 scans (27\%) were reported to have difficult bone structure. The patients of Prosthetic group have major temporomandibular joint (TMJ) disorders and have at least one preoperative scan and one postoperative scan after full prosthetic metallic TMJ-reconstruction with example shown in (Fig.~\ref{fig2}). The patients of Orthognathic surgery group have undergone a bilateral sagittal split osteotomy (BSSO), where the mandible is split bilaterally and moved forward or backward, to correct malocclusion and functionality~\cite{will1989factors}, and have at least one preoperative and one postoperative scan taken six to 12 months after the operation. The BSSO changes the path of the mandibular canal and fixation materials cause metallic artefacts in the imaged area, both seen as differences between the preoperative scan and postoperative scan of the patient. 
The effect of this heterogeneity is shown in three dimensional (3D) reconstructions models of preoperative (Fig.~\ref{fig3}a) and postoperative (Fig.~\ref{fig3}b) cases. The patients of Pathological group have severe pathological bone destructive diseases in the mandible, which may partly or completely destroy the visibility of the mandibular canals, and have at least one preoperative, and one postoperative or follow-up scan. Eight of the 14 patients (57\%) had a malignant bone destruction on the right side of the mandible. 

 The immediate postoperative scans, i.e. those imaged less than five days after the BSSO surgery, were excluded from the main analysis. The  visibility of mandibular canal is extremely poor in these scans~\cite{politis2013visibility} and there are changes to the path of the mandibular canal due to BSSO. Thus, the localisation of the mandibular canal is challenging in these scans. However, the task of localising the mandibular canal in these cases is rarely clinically relevant and thus the analysis is included in the Supplementary.

This study is based on a retrospective and registration dataset and as such does not involve experiments on humans and/or the use of human tissue samples and no patients were imaged for this study. A registration and retrospective study does not need ethical permission or informed consent from subjects according to the law of Finland (Medical Research Act (488/1999) and Act on Secondary Use of Health and Social Data (552/2019)) and according to European General Data Protection Regulation (GDPR) rules 216/679. The use of the Finnish imaging data was accepted by the Tampere University Hospital Research Director, Finland October 1, 2019 (vote number R20558).

\noindent

\subsection*{Quantitative analysis approach}
In the quantitative analysis, the outputs of the DLS were compared to the senior radiologist's annotations. The mandibular canal localisation accuracy of the DLS was evaluated using the Symmetric Mean Curve Distance (SMCD)~\cite{jarnstedt2022comparison}. In the SMCD, the average deviation between the full path of the radiologist's annotation and the DLS output is determined. It thus measures how far apart the two curves are on average. In addition, the segmentation performance of the DLS was evaluated using the Dice similarity coefficient (DSC) and average symmetric surface distance (ASSD) that is a normalised measure of intersection of two segmentations and average error between outer segmentation surfaces, respectively.  However, these metrics are computed using the senior radiologist's approximate segmentation of the canal as the ground truth, thus they only approximate the true segmentation performance.

The reproducibility was evaluated by measuring the temporal variability in SMCD for each canal, i.e.\ the two canals of each patient were treated as independent subjects of interest. The reproducibility measures we used were the within-subject standard deviation (wSD) and the repeatability coefficient (RC), and we also report the within-subject mean and ranges of values. The wSD estimates the variability in SMCD values for the canals and the RC the largest difference between two SMCD values for the same canal with 95\% confidence. These quantities are computed in Eqs.~\eqref{eq:wsd} and \eqref{eq:rc} as follows~\cite{barnhart2009applications}:
\begin{align}
    \text{wSD} &= \sqrt{\text{WMS}}, \label{eq:wsd} \\
    \text{RC} &= 1.96~\sqrt{2}~\text{wSD} \label{eq:rc} \nonumber \\ &\approx 2.77 \text{wSD}, 
\end{align}
where WMS is the within-subject mean squared error, computed as:
\begin{equation*}
    \text{WMS} = \sum_{i=1}^n \sum_{k=1}^{K_i} \frac{(Y_{ik} - \bar{Y}_i)^2}{n(K-1)},
\end{equation*}
$n$ is the number subjects, $K_i$ the number of repetitions for subject $i$, $Y_{ik}$ the $k$th measurement of subject $i$, and $\bar{Y}_i$ the mean of replications of subject $i$. As our dataset includes variable number of repetitions per subject, the $K$ is computed by applying a downward correction to reduce overestimation of the variation among smaller groups compared to larger groups: $K\!=\!\frac{1}{n-1}\left(N-\frac{\sum_{i=1}^n\!K_i^2}{N}\right)$, where $N$ is the total number of scans~\cite{nakagawa2010repeatability}.

\subsection*{Qualitative analysis approach}
For the qualitative analysis, the annotations by the senior radiologist and those produced by the DLS were assessed by three experts; two dental and maxillofacial radiologists with 15 and 13 years of working experience, and one resident of dental and maxillofacial radiology with 8 years of working experience in dentistry. 

There were in total 636 mandibular canal segmentations in the comparison: 318 annotated by the radiologist and 318 DLS outputs, which were provided to the three experts independently and in a random order, without informing whether the segmentation was done by the radiologist or by the DLS. In addition, the segmentations were assessed two times by each expert in separate sessions with a two week interval. The observers assessed the quality of each segmentation and its feasibility for clinical diagnostics using a five-point Likert scale defined with 0: not usable for diagnostics, 1: partly usable for diagnostics, canal visibility below one half, 2: minor issues with almost fully usable for diagnostics, 3: almost perfect, fully diagnostically usable, and 4: perfect, fully diagnostically usable. The Likert ratings were also used to form a binary scale for the diagnostic suitability of a canal segmentation. It was defined as Likert ratings 3 and 4 being fully suitable for diagnosis and 0-2 as not fully suitable.

The observers justified their ratings using five pre-selected error types: fully missing, major parts of the segmentation missing, slightly off the centre of the canal, short at the mandibular foramen, and short at the mental foramen. The first and second error types are considered clinically relevant resulting in Likert rating of 0-2 and diagnostically not fully usable. The other three error types have less clinical importance and they can be explained by the anatomy of the mandible, and the nature of the heterogeneity subgroups. 

The reproducibility was evaluated using the Likert scores of the first annotation session. Due to the discrete and ordinal nature of the Likert scores, as well as the high consistency of the given scores in the dataset, Gaussian random effects models and analysis of variance methods were deemed unsuitable. In addition, the number of temporal scans varied between the patients, and thus methods that require paired data, such as the Kendall rank correlation coefficient, could not be used to analyse the reproducibility. We utilised a Bayesian statistical analysis approach for the reproducibility of ordinal measurements, proposed in a recent study~\cite{bayesordinal}. In short, it uses a random effects model, with subject-dependent random effects, combined with the Master's partial credit model~\cite{partialcredit}. It defines the probability that the subject $i$'s Likert score $y_i$ is equal to the number $h$, given the random effect of the subject $x_i$ and grader $j$ specific parameters $\alpha_j$ and $\delta_j$, as follows:
\begin{equation*}
    p_j(y_i=h \mid x_i, \alpha_j, \delta_j) = \frac{\exp{(\sum_{m=1}^{h-1}\alpha_j(x_i-\delta_{jm}))}}{\sum_{n=1}^H\exp{(\sum_{m=1}^{n-1}\alpha_j(x_i-\delta_{jm}))}}. 
\end{equation*}
It should be noted that in computing the probabilities, the Likert scores were mapped from 0-4 to 1-5. A detailed description of the approach is given in the Supplementary material.

We used the \textit{repeatability measure} (RM) proposed in the aforementioned study to measure the reproducibility. For grader $j$, it is defined as the probability that two Likert scores for the same segmentation will be the same, averaged over the data in Eq.~\eqref{eq:rm}:
\begin{equation}
    \hat{RM}_j(\alpha_j, \delta_j) = \frac{1}{n} \sum_{i=1}^n \sum_{h=1}^5 p_j(y_i=h \mid x_i, \alpha_j, \delta_j)^2 \label{eq:rm}
\end{equation}
The Bayesian analysis produces a distribution of possible parameter values $\alpha_j$ and $\delta_j$, which we used to compute the posterior average $RM_j = Ave_{\alpha_j, \delta_j}(\hat{RM}_j)$ and the 95\% credibility intervals for each Expert. The analysis was performed for the full dataset, as well as for each heterogeneity group independently. We implemented the analysis using the Stan~\cite{stan} programming language with the Python interface PyStan~\cite{pystan} using the default settings described in the original study~\cite{bayesordinal}.

To analyse the intra- and interobserver variability, we calculated the accuracy of Likert ratings.
The interobserver variability was evaluated by comparing the assessments between each pair of experts, and the intraobserver variability by comparing the ratings of the first and second assessment session.

\section*{Results}\label{sec2}

This section presents the results for the quantitative analysis of the DLS performance, and for the qualitative analysis approach for the radiologist's annotations and the DLS performance.

\subsection*{Quantitative analysis}

\begin{figure*}[ht]%
\centering
\includegraphics[width=\textwidth]{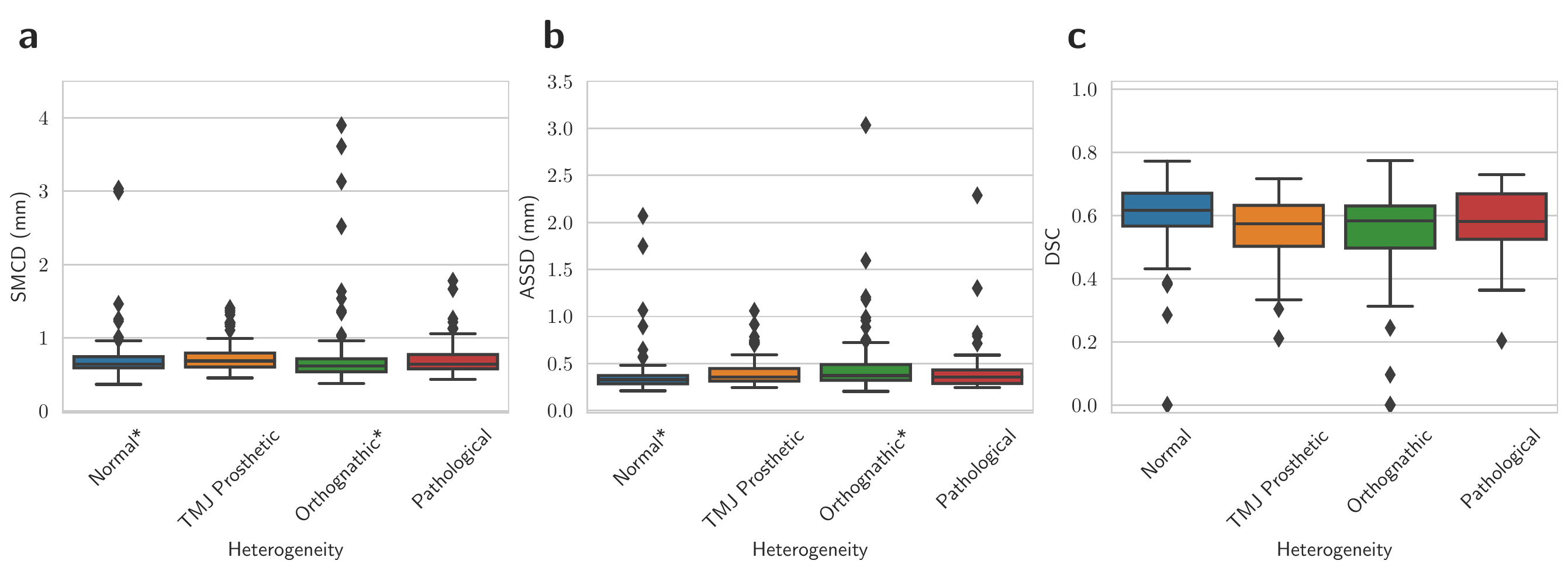}
\caption{Quantitative performance of the DLS for each heterogeneity group. a. Symmetric Mean Curve Distance (mm), b. Average symmetric surface distance (mm), c. Dice similarity coefficient. $^{*)}$two and four values omitted in distance measures from Normal and Orthognathic groups, respectively, as the DLS did not find canals for these cases.}\label{fig4}
\end{figure*}

\begin{figure*}[ht]%
\centering
\includegraphics[width=0.9\textwidth]{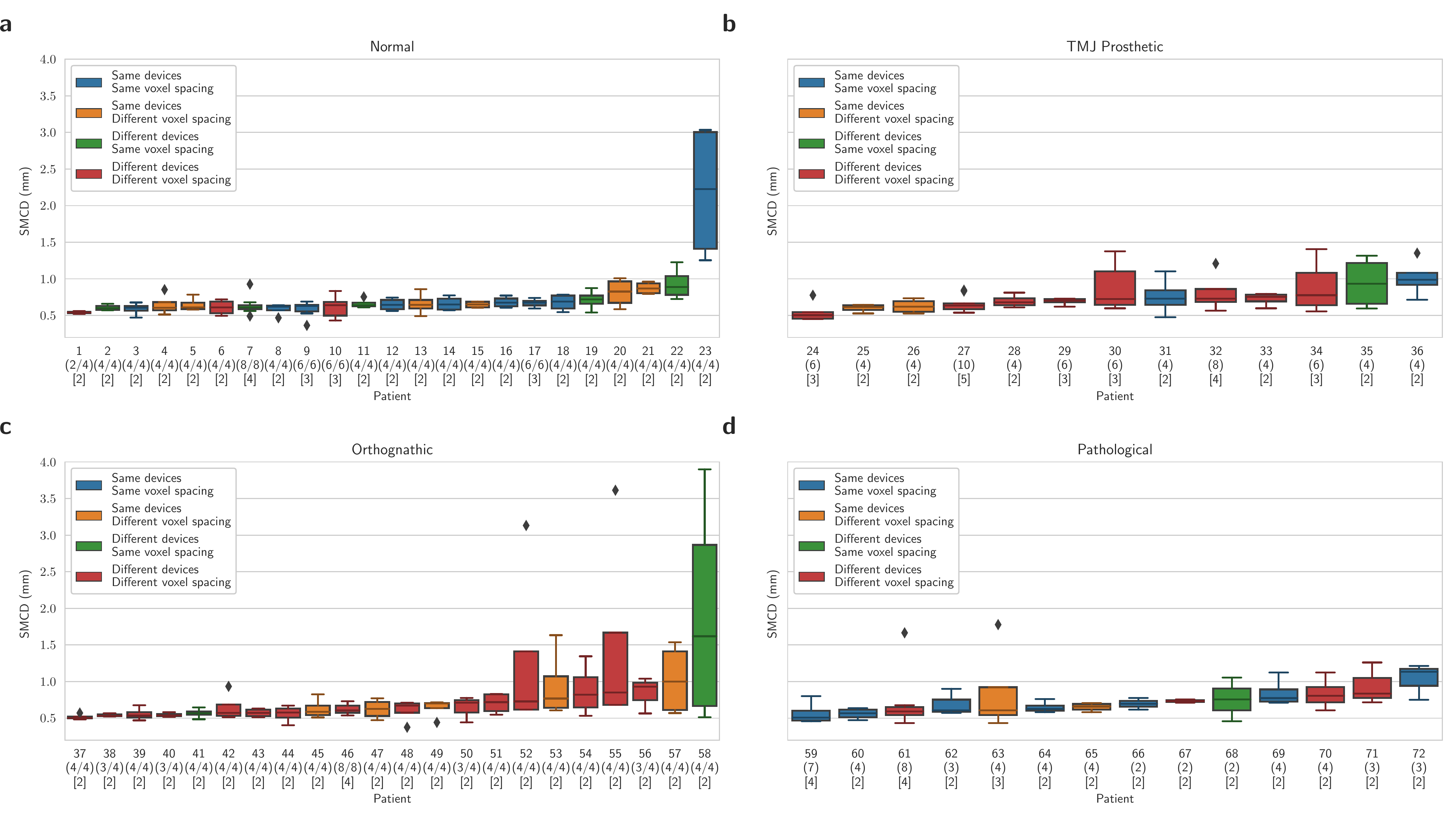}
\caption{DLS performance evaluated with SMCD for each patient. a. Normal, b. TMJ Prosthetic, c. Orthognathic, and d. Pathological heterogeneity group. Parentheses show the number of valid canals out of the total number of canals, and the square brackets the number of scans for each patient. Each patient is colored with the temporal imaging configuration, indicating whether the scans are from the same or different devices, and with the same or different voxel spacings}\label{fig5}
\end{figure*}

The DLS had a repeatability coefficient of 0.969~mm for the full dataset, and when evaluated for each heterogeneity group, the RC was 0.329~mm for Normal, 0.574~mm for TMJ Prosthetic, 1.707~mm for Orthognathic, and 0.648~mm for Pathological group. Full results are shown in \cref{tab2}. The overall localisation performance of the DLS, measured using the median and interquartile range (IQR) of SMCD on valid canals, was found to be 0.643 (0.186)~mm for the full dataset, and 0.638 (0.151)~mm, 0.683 (0.197)~mm, 0.617 (0.178)~mm, and 0.639 (0.201)~mm for Normal, TMJ Prosthetic, Orthognathic, and Pathological heterogeneity groups, respectively. The segmentation performance of DLS measured using the median and IQR of the average symmetric surface distance (ASSD) on valid canals was 0.351 (0.135)~mm for the full dataset, and 0.328 (0.088)~mm, 0.356 (0.133)~mm, 0.365 (0.148)~mm, and 0.355 (0.145)~mm for Normal, TMJ Prosthetic, Orthognathic, and Pathological heterogeneity groups, respectively. The Dice similarity coefficient (DSC) on all the canals was 0.567 (0.133) for the full dataset, and 0.597 (0.121), 0.550 (0.114), 0.540 (0.163), and 0.579 (0.108) for Normal, TMJ Prosthetic, Orthognathic, and Pathological heterogeneity groups, respectively.
Full results of the localisation and segmentation performance of the DLS for each heterogeneity group is shown in Fig.~\ref{fig4}. In addition, the patient level reproducibility is presented in Fig.~\ref{fig5}.

\begin{table}[ht]
\caption{Within-subject mean and standard deviation (wSD), repeatability coefficient (RC), and the range of valid SMCD (mm) values. The results are presented for the full dataset and for each heterogeneity group separately. N~=~the total number of canals, n~=~the number of unique canals, and K~=~the set of the number of canals present in the group.
}\label{tab2}%
\centering
\begin{tabular}{@{}llllllll@{}}
Heterogeneity & n & N & K & Mean & wSD & RC & Range\\ 
\hline
Full dataset & 131 & 302 & \{2, 3, 4, 5\} & 0.761 & 0.350 & 0.969 & [0.471, 3.014] \\
Normal & 44 & 98 & \{2, 3, 4\} & 0.743 & 0.119 & 0.329 & [0.554, 3.014] \\
TMJ Prosthetic & 26 & 70 & \{2, 3, 4, 5\} & 0.756 & 0.207 & 0.574 & [0.487, 1.172] \\
Orthognathic & 40 & 84 & \{2, 4\} & 0.806 & 0.616 & 1.707 & [0.471, 2.206] \\ 
Pathological & 21 & 50 & \{2, 3, 4\} & 0.719 & 0.234 & 0.648 & [0.515, 1.049] \\
\end{tabular}
\end{table}

\subsection*{Qualitative analysis}

In the qualitative analysis, the reproducibility of the Likert rating for the radiologist and the DLS, measured in the repeatability measure, was found to be 0.958 and 0.895 for Normal, 0.841 and 0.887 for TMJ Prosthetic, 0.945 and 0.913 for Orthognathic, and 0.916 and 0.886 for Pathological groups, respectively. In terms of diagnostically fully usable canal paths, the RM values for the radiologist and the DLS were 0.992 and 0.954 for Normal, 0.963 and 0.974 for TMJ Prosthetic, 0.978 and 0.922 for Orthognathic, and 0.997 and 0.980 for Pathological groups, respectively. Full results are shown in \cref{tab1}.

\begin{table}[ht]
\caption{Qualitative reproducibility results, in terms of the repeatability measure, for the Likert scoring of the radiologist's annotation and the DLS output. Results shown for each heterogeneity group and radiologist, as well as for the full dataset and average of the Experts. CI denotes the Bayesian credibility interval.
}\label{tab1}%
\centering
\begin{tabular}{@{}llll@{}}
Reproducibility &  Heterogeneity & Radiologist Mean (95\% CI) & DLS Mean (95\% CI)\\ 
\hline
\multirow{5}{*}{Average}  
  &  Full dataset  & \textbf{0.923 (0.842, 0.979)} & 0.877 (0.814, 0.932)\\
  &  Normal  & \textbf{0.958 (0.882, 1.000)} & 0.895 (0.833, 0.935)\\
  &  TMJ Prosthetic  & 0.841 (0.725, 0.925) & \textbf{0.911 (0.829, 0.970)}\\ 
  &  Orthognathic  & \textbf{0.945 (0.844, 1.000)} & 0.815 (0.711, 0.907)\\
  &  Pathological  & \textbf{0.916 (0.747, 0.996)} & 0.899 (0.712, 0.998)\\\hline
  
\multirow{5}{*}{Expert 1}  
  &  Full dataset  & \textbf{0.964 (0.937, 0.983)} & 0.915 (0.887, 0.936)\\
  &  Normal  & \textbf{0.996 (0.976, 1.000)} & 0.913 (0.864, 0.939)\\
  &  TMJ Prosthetic  & 0.865 (0.776, 0.933) & \textbf{0.936 (0.872, 0.977)}\\ 
  &  Orthognathic  & \textbf{0.995 (0.970, 1.000)} & 0.866 (0.796, 0.917)\\
  &  Pathological  & 0.960 (0.881, 0.997) & \textbf{0.969 (0.905, 1.000)}\\\hline
  
\multirow{5}{*}{Expert 2}  
  &  Full dataset  & \textbf{0.932 (0.899, 0.958)} & 0.874 (0.838, 0.904)\\
  &  Normal  & \textbf{0.950 (0.893, 0.984)} & 0.886 (0.828, 0.925)\\
  &  TMJ Prosthetic  & 0.856 (0.766, 0.927) & \textbf{0.913 (0.841, 0.961)}\\ 
  &  Orthognathic  & \textbf{0.947 (0.888, 0.982)} & 0.793 (0.704, 0.861)\\
  &  Pathological  & \textbf{0.959 (0.882, 0.997)} & 0.928 (0.842, 0.981)\\\hline
  
\multirow{5}{*}{Expert 3} 
  &  Full dataset  & \textbf{0.872 (0.829, 0.907)} & 0.841 (0.801, 0.875)\\
  &  Normal  & \textbf{0.928 (0.864, 0.970)} & 0.887 (0.827, 0.925)\\
  &  TMJ Prosthetic  & 0.800 (0.697, 0.891) & \textbf{0.885 (0.810, 0.934)}\\ 
  &  Orthognathic  & \textbf{0.893 (0.821, 0.944)} & 0.786 (0.698, 0.856)\\
  &  Pathological  & \textbf{0.829 (0.709, 0.918)} & 0.798 (0.678, 0.885)\\
\end{tabular}
\end{table}

\noindent

In terms of descriptive statistics, the two ratings of the three experts resulted in 1908 ratings, both for the radiologist and the DLS. 1893 (99.2\%) canal paths from the radiologist and 1828 (95.8\%) from the DLS were rated diagnostically fully usable. Taking the median rating of the Experts for each canal showed that 316 (99.4\%) canals by the radiologist and 305 (95.9\%) canals by the DLS were fully suitable for diagnosis, with 1 (0.3\%) canal for the radiologist and 2 (0.6\%) canals for the DLS uncertain, and with 1 (0.3\%) canal for the radiologist and 11 (3.5\%) canals for the DLS not fully diagnostically usable. The mean, standard deviation, and counts of the Likert ratings for each heterogeneity group are shown in \cref{tab3}.

\begin{table}[!ht]
\caption{Mean, standard deviation (SD), and counts of all the Likert ratings ($N_L$ for the Likert score $L$) grouped by heterogeneity and the marker (the radiologist or the DLS).
}\label{tab3}%
\centering
\begin{tabular}{@{}lllllllll@{}}
 Heterogeneity & Marker & Mean & SD & $N_0$ & $N_1$ & $N_2$ & $N_3$ & $N_4$  \\ \hline
\multirow{2}{*}{Full dataset}  &  Radiologist  &  \textbf{3.94}  &  \textbf{0.27}  & 0 & 3 & 12 & 72 & 1821 \\ 
                               &  DLS  & 3.84 & 0.65 & 35 & 9 & 35 & 75 & 1753 \\ \hline
\multirow{2}{*}{Normal}        &  Radiologist &  \textbf{3.97} &  \textbf{0.19}  & 0 & 0 & 2 & 15 & 595 \\
                               &  DLS & 3.86 & 0.62 & 12 & 0 & 8 & 21 & 571 \\ \hline
\multirow{2}{*}{TMJ Prosthetic}   &  Radiologist & 3.88 & 0.42 & 0 & 3 & 6 & 28 & 383 \\
                                  &  DLS &  \textbf{3.92} &  \textbf{0.38}   & 0 & 4 & 2 & 18 & 396 \\ \hline
\multirow{2}{*}{Orthognathic}    &  Radiologist  &  \textbf{3.97} &  \textbf{0.22}  & 0 & 0 & 4 & 11 & 537 \\
                                  &  DLS & 3.71 & 0.90 & 23 & 3 & 21 & 17 & 487 \\ \hline
\multirow{2}{*}{Pathological}    &  Radiologist  &  \textbf{3.94} &  \textbf{0.23} & 0 & 0 & 0 & 18 & 306 \\
                                  &  DLS  & 3.90 & 0.39 & 0 & 2 & 4 & 19 & 299 \\ 
\end{tabular}
\end{table}

The error types reported by the Experts were also aggregated by majority voting for each of the canals. The radiologist's annotations had errors for cases in TMJ Prosthetic group, specifically, 2~(2.9\%) were short at mandibular foramen and 1~(1.4\%) slightly off centre. The errors of the DLS were; fully missing canal for 2~(2.0\%) cases from the Normal and 4 (4.3\%) from the Orthognathic group, major parts missing in 1~(1.4\%) case from TMJ Prosthetic group, short at mandibular foramen for 1~(1.5\%) case from Pathological and 1~(1.4\%) case from TMJ Prosthetic, and short at mental foramen for 2~(2.0\%) from Normal and 2~(2.2\%) from Orthognathic group.

In terms of the interobserver agreement of the Likert rating, the accuracy between Expert 1 and Expert 2 was found to be 94.81\% and 95.91\%, between Expert 1 and Expert 3 to be 91.98\% and 91.51\%, and between Expert 2 and Expert 3 to be 91.04\% and 90.87\%. The intraobserver agreement was found to be 99.37\%, 95.91\%, and 92.45\% for Experts 1, 2 and 3, respectively.

\section*{Discussion}\label{sec12}

In this study, we have demonstrated the reproducibility capacity of a deep learning -based automatic mandibular canal localisation system on a heterogeneous set of temporal clinical data. The DLS achieved both a high localisation performance and reproducibility in our systematic quantitative and clinical performance evaluation of multiple graders. In addition, the present work is based on an out-of-distribution dataset that highlights the evidence of the temporal generalisation capacity of the DLS.

In the quantitative analysis, from the point of view of clinical usability the DLS produced good repeatability coefficient values for Normal, TMJ Prosthetic, and Pathological groups. The Orthognathic group had the worst mean, standard deviation, and repeatability coefficient values, which can be explained by the canals being affected by bilateral sagittal split osteotomy, which were generally difficult for the DLS. In this operation, the osteotomy lines go into the posterior mandibular canal areas, and after the ossification in the postoperative CBCT scans, the shapes, paths, and lengths of the mandibular canals may become modified. The postoperative mandibular canal can even have accessory openings to the surface of the mandible, and it can be asymmetrical to the contra-lateral mandibular canal. The generalisation performance for the BSSO cases is worse likely due to the lack of these cases in the training data. As for the overall generalisability results for SMCD, ASSD, and DSC, the DLS shows similar characteristics  with the previous studies for the Normal, TMJ Prosthetic, and Pathological heterogeneity groups~\cite{jaskari2020deep, jarnstedt2022comparison}. Specifically, the median SMCD values obtained for the DLS were lower than reported interobserver variability~\cite{jarnstedt2022comparison}. Additionally, the mean DSC was within the range of interobserver variability of 0.46--0.60 reported in the Supplementary section of that study. Furthermore, the DLS had large localisation and segmentation errors only on few canals that can be seen as outliers in the result figures. There is a  similar proportion of outliers as in the previous study~\cite{jarnstedt2022comparison} and they are likely to be caused by some artefacts or difficult heterogeneity.

In the qualitative or clinical analysis, the radiologist had a higher probability for receiving the same Likert grading on two temporal segmentations, i.e.\ the repeatability measure, than the DLS on Normal and Orthognathic groups. On the Pathological group, the radiologist had a higher repeatability measure than the DLS with Expert 2, Expert 3, and on average, but lower repeatability with Expert 1. However, the DLS showed higher repeatability measure on the TMJ Prosthetic group for all the Experts and on average. Challenges in the TMJ Prosthetic group for the radiologist can be caused by metallic artefacts that affect the visibility of the canal (Fig.~\ref{fig2}b). The patients of the group had been operated using an individually designed full TMJ-reconstruction, with condylar component that consists of titanium, and fossa component of Ultra-high molecular weight polyethylene, both fixed to the bone with titanium screws. These metallic structures cause major artefacts on the CBCT scans on the side with the operated mandible, degrading the visibility of the mandibular canal.

Overall, the DLS showed slightly worse performance than the radiologist in terms of fully diagnostically usable canals and when evaluating the average Likert score. The DLS produced different types of errors, such as fully missing canals or major parts missing, while the radiologist's errors were mostly too short or slightly off centre canals. The error of too short canal can be expected, as the length of a mandibular canal is challenging to evaluate in the foramen mandibulae area. This is because of the anatomical features and subjective ending point of the canals~\cite{jarnstedt2022comparison}. The error of a canal being slightly off centre can be caused by an interpolation artefact or human error, when marking the canal manually~\cite{jarnstedt2022comparison}. There were small Likert score differences between the heterogeneity groups for both the radiologist's annotations and the DLS outputs. The group with the worst performance for the radiologist was TMJ Prosthetic group with the average Likert score of 3.88, while the worst performing group for the DLS was Orthognathic group with the average Likert score of 3.70. We found that there were no major differences between the majority of the qualitative results for the different heterogeneity groups. Indeed, most of the differences were caused by a few outlier cases. In terms of the patient level performance, the DLS turned out to have low variation between each scan of a patient, even for most of the cases that were imaged with different scanning devices and voxel spacing. Lastly, we evaluated the intra- and interobserver variability of the Likert scores given by the three experts in the two scoring sessions. We found out that there is a very strong overall intra- and interobserver reliability, but there are some differences between the Experts. Lower interobserver results can be explained by the subjective nature of the mandibular canal annotations, which is based on the radiologist's experience in the task~\cite{hamid2021diameter}.

There are limitations in this study. First, we did not report all of the possible heterogeneities affecting the scans, such as difficult bone structure and movement of the patient. Second, the mandibular canal regions may also be affected by normal dental treatment during the follow-up period, which can cause changes to the nearby areas, such as new implants, dental crowns, endodontic materials, and orthodontic fixed appliances. These should be taken into consideration when designing future follow-up studies. In addition, further development is required to improve the robustness of the system for most anatomically, pathologically, or surgically difficult cases.

This work demonstrates the necessary capabilities of deep learning for longitudinal follow-up studies with different scanners or imaging parameters, and when there are anatomical or pathological changes in the patient’s mandible. The next step in terms of a clinical validation will be the evaluation of the deep learning system under radiologist's supervision. The DLS could be used as an augmenting tool in routine radiological, surgical and, dental practice, which can minimize the potential complications in surgery.

\section*{Conclusion}\label{sec13}

The reproducibility of mandibular canal localisation with a deep learning system was evaluated on a heterogeneous dataset of temporal cone beam computed tomography scans. The reproducibility was found to be comparable to a radiologist in terms of quantitative and qualitative measures. In case of localisation and segmentation, the generalisation performance was found to be within or better than the interobserver variability suggested in the literature. 

\bibliography{bibliography}

\section*{Acknowledgements}

Expert Likert rating and error type reporting used in this study were partly provided by Antti Lehtinen, DDS Specialists of Dentomaxillofacial Radiology and Maarit Jordan DDS Resident of Dentomaxillofacial Radiology.

\section*{Author contributions statement}

J.Jä.\ conception, design of the work, data and annotation acquisition, interpretation of data, data annotation, wrote the main manuscript text; J.S.\ \& J.Ja.\ conception, design of the work, ran the experiments, analysis, wrote the main manuscript text and produced figures; K.K.\ conception, interpretation of data, project management, corresponding author, wrote and polished the main manuscript text; H.M.\ interpretation of data and data annotation; A.H.\ project management, design of the work, analysis; O.S., V.V., V.M.\ project management and design of the work; S.P., S.N.\ project management, design of the work, data acquisition; All authors reviewed the manuscript.

\section*{Data availability}

The datasets used in model training, validation, and testing were provided by TAUH and CMU, and as such is not publicly available and restrictions apply to their use according to the Finnish law and General Data Protection Regulation (EU) and to the Thai law, respectively.

\section*{Code availability}

The code used for pre- and postprocessing and the deep learning techniques includes proprietary parts and cannot be released publicly. However, the approach can be replicated using the information from the cited literature and the evaluation measures with the equations in the “Methods” section and in Supplementary Information.

\section*{Additional information}

The authors declare no competing interests.

\end{document}